# Testing the trade-off between productivity and quality in research activities[1]

Giovanni Abramo[*]

*Italian Research Council and Laboratory for Studies of Research and Technology Transfer at*

*University of Rome "Tor Vergata" – Italy*

ADDRESS: Dipartimento di Ingegneria dell'Impresa, Università degli Studi di Roma "Tor Vergata",

Via del Politecnico 1, 00133 Roma - ITALY, tel. and fax +39 06 72597362, abramo@disp.uniroma2.it

Ciriaco Andrea D'Angelo

*Laboratory for Studies of Research and Technology Transfer at*

*University of Rome "Tor Vergata" – Italy*

ADDRESS: Dipartimento di Ingegneria dell'Impresa, Università degli Studi di Roma "Tor Vergata", Via

del Politecnico 1, 00133 Roma - ITALY, tel. and fax +39 06 72597362, dangelo@disp.uniroma2.it

Flavia Di Costa

*Laboratory for Studies of Research and Technology Transfer at*

*University of Rome "Tor Vergata" – Italy*

ADDRESS: Centro Interdipartimentale "Vito Volterra", Università degli Studi di Roma "Tor Vergata",

Via Columbia 2, 00133 Roma - ITALY, tel. +39 06 72595400, dicosta@disp.uniroma2.it

---

[*] **Corresponding author**

**Abstract**

In recent years there has been an increasingly pressing need for the evaluation of results from public sector research activity, particularly to permit the efficient allocation of ever scarcer resources. Many of the studies and evaluation exercises that have been conducted at the national and international level emphasize the quality dimension of research output, while neglecting that of productivity. This work is intended to test for the possible existence of correlation between quantity and quality of scientific production and determine whether the most productive researchers are also those that achieve results that are qualitatively better than those of their colleagues. The analysis proposed refers to the entire Italian university system and is based on the observation of production in the hard sciences by above 26,000 researchers in the period 2001 to 2005. The results show that the output of more productive researchers is superior in quality than that of less productive researchers. The relation between productivity and quality results as largely insensitive to the types of indicators or the test methods applied and also seems to differ little among the various disciplines examined.

# 1. Introduction

In recent years, the debate over the evaluation of publicly funded research has attracted interest from a wide and varied public. More evaluation exercises and analyses are being implemented, both for the examination of public institutions and individual researchers. Comparative evaluation would allow stimulation of greater productive efficiency, permit allocation of resources in function of merit and reduce information asymmetry between demand and supply of knowledge.

However, comparative evaluation of research activity results as quite complex, especially along the dimension for labor productivity. The national research assessments implemented in various nations have so far taken a peer review approach, avoiding the evaluation of research productivity and all of its associated difficulties. These assessments emphasize quality in research output, which is more readily measurable, and limit observations to a sample of total output provided by each institution under observation (for example 50% in the case of the United Kingdom (RAE, 2008) and 10% in the case of the VTR, Italy (CIVR, 2006). However, national level measures of productivity are not possible with a peer-review approach unless the evaluation is extended to the entire scientific production of the institutions being evaluated. This would result in exercises with costs and times of execution so high as to discourage their actual implementation. Meanwhile, basing measures of productivity on data provided by the actual research institutions also engenders serious risks, as demonstrated by the only experience of this kind, in Australia (Composite Index)[2].

---

[2] Audits conducted by KPMG on publication lists submitted by universities found a high error rate (34% in 1997). 97% of errors affected final scores, and consequently funding allocations (Harman, 2000).

The question remains open as to whether labor productivity and the quality of product proceed in step, or whether there is instead a trade-off between the two. Is it the case that the most productive scientists also achieve the highest quality of results, or is it the case that higher levels of production detract from quality? Answers could prove useful both in the phases of formulating evaluation exercises and in the development of associated reward systems.

The literature that might provide answers to this intriguing question seems exceptionally scarce, and the reason is quickly evident. While comparative measures of quality are quite readily achieved (Hicks, 2009; RAE, 2008; CIVR, 2006; Van Raan, 2005; VSNU, 2002; NAS-NAE, 1999), either with peer review methods (as in the national evaluation exercises) or through a bibliometric approach, it is as much more complex to measure and compare research productivity. Here the literature reports few analyses, and all are limited to a few scientific disciplines or a restricted number of institutions (Macri and Dipendra, 2006; Kalaitzidakis et al., 2003; Pomfret and Wang, 2003). None of the authors of these analyses went so far as the step of actually investigating for correlation of productivity and quality of research.

The production of new knowledge is a function of multi-input/output type, which means that the problem of measuring productivity is also multi-faceted. For output, there are multiple modes of codifying the production of new knowledge, adopted to varying extent by different disciplines. Fertility of scientific publication is different from discipline to discipline, as is the degree of coverage by international bibliometric databases, such as the Thomson Reuters Web of Science (WoS), or Elsevier Scopus. These databases provide a reliable reference for the hard sciences, where publication is by far the most widely used form for codification of research output. However the coverage and representativity of these databases is less for disciplines in which scientists favor publication in national journals, such as in certain branches of social sciences and jurisprudence, or where other forms of codification dominate, such as in the arts and humanities[3].

Though the representativity of the databases is accepted concerning the hard sciences, two very substantial obstacles have deterred bibliometrists from proceeding to comparative measures at the national level. The first concerns the identification and reconciliation of the different ways in which the same organization's name is reported in the "author address" field of the databases, while another concerns the correct attribution of authorship. This second type of problem originates from the fact that the bibliometric records lack detailed links between the individual authors listed and the list of their home institutions, and, at the outset, from the way in which the author names are reported in the databases (only the initial of the first name, for the WoS). Because of these shortcomings, in cases of co-authorship, it had not always been possible to unequivocally associate the author's name with his or her institution. However, these technical limitations have been addressed and overcome by (Abramo et al., 2008a), who develop an effective disambiguation procedure to attribute listed publications to unequivocally identified university authors.

Difficulties for equitable comparison of research productivity also arise on the input side of the production function: factors of production, with the possible exception of labor, are not always easily measurable, nor can they be attributed with certainty to individual productive units. Indeed, even labor is difficult to express in hours, since the portion of time dedicated to research is different between scientists in universities and research institutes. Capital and other factors that have a direct or indirect impact on productivity are equally difficult to measure and attribute to individual productive units. This includes factors such as geographic location[4], the accumulated knowledge of the home institution, etc. Therefore a subsequent normalization with respect to these variables is not possible. Even the quality of research output, as measured through national peer review exercises, could be influenced by these variables[5]. However, none of these assessments have included normalization in their methodologies.

---

[3] The Italian research assessment exercise provides an example of the phenomenon. Here, the products that the universities submitted for evaluation were included in the WoS in over 90% of the cases, for hard sciences, but only 72% of cases for economics and statistics and only 15% of cases for the other social sciences, the arts and humanities.

Given this discussion, it is not surprising that a literature search by the authors found only one study concerning the research question at issue. Costas et al. (2009) analyze the relationship between productivity and quality of output for 1,038 researchers working at the Spanish Council for Scientific Research, in three areas: biology and biomedicine, material science and natural resources. The results show that the total number of citations received by the researchers, for the period 1994-2004, increases in a cumulatively advantageous way[6] in function of the number of publications. The cumulative advantage is greater for researchers who publish in low citation-density fields compared to those who publish in high citation-density fields.

Preceding studies concerning size-dependent cumulative advantage referred to aggregate research units, either research groups (Van Raan, 2006a), institutes or countries (Katz, 2000; Katz, 2005), which prevents examination of correlation between researcher productivity and output quality.

---

[4] Due to the geographic proximity effect, locations of universities in areas with high intensity of public and private research can favor scientific collaboration and greater research productivity (Abramo et al., 2009a).

[5] Abramo et al. (2009a) demonstrate that publications in co-authorship with other organizations have an average quality that is greater than for those produced "in house". Since location can have an impact on opportunities for collaboration, it can thus also have an effect on quality of output.

[6] "By *cumulative advantage* we mean that the dependent variable (for instance, number of citations of a group) increases in a disproportional, nonlinear (in this case: power law) way as a function of the independent variable" (Van Raan, 2006a).

The present work aims to provide a robust response to the initial research question, through an approach in which investigation is not limited to a single institution or a few scientific disciplines. The study observations concern all Italian universities (82) and all the hard sciences, for a total of more than 26,000 research staff in 165 disciplinary sectors. The analysis will evaluate the scientific production indexed by the WoS for the period from 2001 to 2005, consisting of a total of over 124,000 publications. The investigation will be conducted at the level of individual disciplinary sectors in order to reduce distortions related to aggregate measurement. This will also permit recognition of variations in the degree of correlation between quality and productivity in function of discipline or disciplinary sector.

## 2. Methodology

The approach adopted in this work is completely bibliometric. This implies the selection of i) indicators for measurement of the quality of scientific output and the productivity of research activity; ii) data sources; iii) the methods to determine if more productive scientists also produce results of higher quality. This section addresses the selection of indicators and data sources, while the first part of Section 4 deals with the third issue.

*2.1 Indicators of measure*

A first question concerning methodology is the definition of the research output that will be measured. As noted, there are multiple forms of codification for new knowledge produced by research activity. Having limited the field of analysis to the hard sciences, the choice of scientific publications as a proxy for research output certainly finds support in the literature (Moed et al., 2004). While it may now be standard practice to use citations as a proxy of the quality of a research product, it should be noted that the number of citations actually measures the quantity of response to a research work, but does not necessarily constitute a judgment of "goodness" or validity (Moed and Hesselink, 1996). Still, the literature includes various studies that show existence of a positive relationship between citations to a work and the experts' opinions concerning the quality and importance of the same work (Abramo et al., 2009b; Van Raan, 2006b; Aksnes and Taxt, 2004; Rinia et al., 1998; Moed and Hesselink 1996). This leads us to use citations as the proxy for the quality of a publication. Since the rate of citations is particularly sensible to the discipline to which a publication belongs, we have conducted the analysis by ISI subject category[7] and defined the normalized quality index for publications ($QI_c$):

$QI_c^{ave}$ = Number of citations of an article divided by the average number of citations of all articles of the same year and the same ISI category. For instance, a value of 1.40 indicates that the article was cited 40% more often than the average.

---

[7] The ISI subject categories are the scientific disciplines that the WoS uses for the classification of publications.

Since the distribution of citations generally results as typically very skewed in all disciplines, it was seen as opportune to also consider another method of normalization for the citations, the percentile. Here, the quality index, $QI_c^{perc}$ will be:

$QI_c^{perc}$ = ranking of a publication, measured on a 0 – 100 scale, according to the citation distribution of publications of the same year and the same ISI category. A value of 90 indicates that 90% of the articles of the same year in the same ISI category have a lower number of citations than the one under observation.

Like the rates of citations, the life cycles of citations also differ among disciplines. There are noticeable fluctuations in the "cited half-life indicator" across disciplines (Van Raan, 2004; Redner, 2005). In some disciplines, such as mathematics, the intensity of citations increases very slowly over time. In this work, since the time-point for observing citations (January 2008) is quite close to the end of the range of observations (2001-2005), the adoption of citations as proxy of quality might not be completely trustworthy for some disciplines. To give further robustness to the results of the examination we decided to also use the impact factor of the journal as a proxy for quality of individual publications. The literature does raise cautions about such an operation (Moed and Van Leeuwen, 1996; Weingart, 2005). However, the impact factor is certainly an indicator of the prestige of a journal and thus, though it may be difficult to accept it as proxy of the quality of a publication, it permits dealing with the question: do more productive scientists publish in journals that are more prestigious than those in which their less productive colleagues publish? We will thus have the quality index of a specific research publication, measured through the impact factor of the relative journal ($QI_{if}$):

$QI_{if}$ = impact factor ranking of the journal, measured on a 0 – 100 percentile scale according to the impact factor distribution of the journals that publish papers in the same ISI category. A value of 90 indicates that 90% of publications falling in the same category are in journals with lower impact factor than the one under observation.

For measurement of labor productivity, we have formulated two indicators:

- Productivity (P): total of publications authored by a scientist in the period under observation;

- Fractional Productivity (FP): total of the contributions to publications authored by a scientist, with "contribution" defined as the reciprocal of the number of co-authors of each publication.

The research productivity of individual scientists will not be normalized for actual hours worked on research, other productive factors or intangible resources, due to the complete lack of data that can be attributed to individuals. These factors should impact on both quality and quantity of output in a similar and consistent manner and their omission should not jeopardize the results of the analysis.

*2.2 Data, data sources and field of observation*

The data used in the study are taken from the Observatory on Public Research in Italy (ORP), a bibliometric database derived by the authors from the WoS. The ORP provides a census of scientific production from all research institutions situated in Italy. Beginning from this database, the next step was to extract the publications authored by Italian universities in the period 2001-2005, which amounted to a total of roughly 147,000. Through the development of a complex algorithm for recognition of addresses and disambiguation of the real identity of the authors, it was possible to attribute each publication to the responsible university scientists. In large populations of scientists the rate of homonyms among names is very high: 12% of the 60,000 scientists in the Italian university system have names that are homonyms of those for other scientists. Eliminating ambiguities as to the precise identity of the author within acceptable margins of error is a daunting task, explaining why bibliometric studies are generally carried out at aggregated level of analysis, such as at the university level. In the past, analyses conducted at the level of single scientists or research groups were generally limited to a maximum of a few organizations or scientific disciplines, in which case it is possible to disambiguate manually. However, for the 147,000 Italian academic publications indexed in the ORP between 2001 and 2005, the harmonic mean of precision and recall (F-measure) of authorships disambiguated by our algorithm is around 95% (2% margin of error, 98% confidence interval).

Italian regulations require that each university scientist must belong to a specific "Scientific Disciplinary Sector", or SDS. Each sector is in turn part of a "University Disciplinary Area", or UDA. The hard sciences consist of nine UDA[8] and 205 SDS. The correlation analysis for productivity and output quality was conducted for each hard science SDS (165 in total) where at least 50% of the scientists belonging to the SDS had published at least one scientific article in the period under examination.

In the period under consideration, there were 26,273 scientists on staff in the 165 SDS[9] considered. These scientists were identified from the CINECA database of the

---

[8] Mathematics and computer sciences; physics; chemistry; earth sciences; biology; medicine; agricultural and veterinary sciences; civil engineering and architecture; industrial and information engineering.
[9] For more reliable indication of the phenomenon under examination, the analysis excludes all scientists who changed university or SDS or who entered or left the university system during the period of

Italian Ministry of Universities and Research[10].

To give an idea of the barriers overcome and the scope of our field of observation, we report the comments by (Van Raan, 2008), a leading scholar in the field of bibliometrics, concerning his examination of a dataset of 18,000 WoS publication listings by all chemistry researchers in 10 Dutch universities: "This material is quite unique. To our knowledge, no such compilations of very accurately verified publication sets on a large scale are used for statistical analysis of the characteristics of the indicators at the research group level".

**3. Results**

Three distinct analyses were conducted to identify the relation between quantity and quality of scientific production in each SDS.

First, assuming the case that the relation between citations and publications follows a power function law, a regression model was used to test for and quantify the potential existence of increasing returns to scale, or the situation that the citations received by a researcher increase more than proportionally with respect to the publications achieved. Since this test is based on an inferential approach, a further analysis was conducted for the comparison of two distinct subpopulations of scientists selected on the basis of productivity. The objective was to test if the publications of the top 10% of scientists for productivity present, on average, superior quality than the rest of the population, and to what extent. However, this second analysis could also lead to a doubt about whether the

---
observation.
[10] http://cercauniversita.cineca.it/php5/docenti/cerca.php

comparison between the output quality of the two subpopulations should be conducted for the entire publication production of individual researchers or for a selection of the best products of each researcher. One could suppose that some scientists would prefer to produce less output, but of a higher quality. Therefore a third analysis was conducted to consider the best publication of each researcher, and using both the methodologies described, examine the question of whether its quality is correlated (or not) to the productivity of the researcher.

The convergence of results from the three analyses gives significant evidence of the presence and nature of the link between productivity and quality of output. For purposes of policy development, these results can be useful both in the phase of elaborating guidelines for evaluation processes and in the successive phases of formulating incentives to favor improvement in individual performance and the aggregate performance of departments, institutions, etc. In order to detect different characteristics with potential policy implications, all the analyses were conducted not only at the general level, but also the individual disciplinary and sectorial levels.

*3.1 Regression analysis*

All researchers that had not received any citations were excluded from the initial dataset. A regression analysis was conducted for the remaining 20,450 scientists, correlating the normalized total of citations[11] with the total number of publications achieved by each researcher. For the citations, the $QI_c^{ave}$ was considered. The dependent variable of the analysis is thus represented as the sum of the values of this indicator, called $C_o$, for all the publications by a single scientist, while the independent variable is represented by the scientist's productivity, P. The objective is to quantify the γ coefficient of a power-law function of type $C_o = aP^γ$. The results for each UDA are shown in [Table 1].

---

[11] The use of $QI_c^{ave}$ rather than a simple count of citations permits limitation of distorting effects related to the different intensity of citation among the subject categories in which individual researchers may have published.

[Table 1]

The power-law function exhibits increasing returns, since the coefficient γ, in addition to being statistically significant, also results as greater than 1 for all UDA. Since γ represents the percentage increase in total citations when total output increases by 1%, the average quality of scientific production by a researcher increases with his or her productivity. The phenomenon seems most prominent for the chemistry and physics disciplines (γ respectively equal to 1.275 and 1.266) and slightly less evident in the mathematics and computer sciences discipline (1.169).

*3.2 Top scientists versus the rest of the population*

We now ask whether the cumulative advantage seen from the preceding analysis also appears in comparisons between populations that are different for their productivity. Researchers without publications were excluded from the original dataset. The remaining 21,505 scientists were subdivided into two groups on the basis of productivity. Specifically, the top 10% of scientists were isolated from the rest of the population: the top scientists are those who, for bibliometric performance under all indicators of productivity, placed in the top 10% of national ranking for their SDS. The objective was to test if the publications of "top scientists" give indexes of quality that are superior to the rest of the population. The mean difference of % rank between top scientists and the rest of the population, for each indicator of quality, is presented in Table 2. The use of % rank is necessary in order to conduct comparisons at an aggregate level: scientific "fertility" varies among the different hard science subject categories and, as a result, so does the distribution of the bibliometric indicators. The % rank permits application of the same scale (0-100, with 100 as best value) to the bibliometric performance of researchers employed in different SDS[12] and thus permits a robust form of comparison of performance between scientists in different disciplines (Abramo et al., 2008b). The results of the analysis are unequivocal: whatever the criteria for selection of top scientists or indicator of quality, there is a considerable mean difference in quality in favor of the top scientists. For example, the top scientists (identified on the basis of P) have a % rank that is higher by 15.4 for $QI_c^{ave}$, by almost 21 for $QI_c^{perc}$ and by 7.5 for $QI_{if}$, compared to the rest of the population.

[Table 2]

To test for possible variation at the level of disciplines, the difference between top scientists and the rest of the population was measured for single UDA. Table 3 presents the comparison between top scientists (identified in terms of P) and the rest of the population: the mean difference between the two sets is measured for each UDA and for each of the three indicators of quality in terms of difference in % rank.

---

[12] For example, a researcher will have a % rank for productivity of 80 if 20% of colleagues in the same SDS register a greater productivity. The same holds for the other bibliometric indicators.

[Table 3]

The data do not seem to indicate a significant variance among disciplines, especially for the indicators of quality based on citations. In terms of $QI_c^{ave}$, the difference ranges from a minimum of 13.4 for earth sciences to a maximum of 19.3 for agricultural and veterinary sciences. For $QI_c^{perc}$, the minimum and maximum differences respectively concern chemistry (18.4) and civil engineering and architecture (26.2). In terms of $QI_{if}$, the maximum difference in performance between top scientists and the rest of the population is seen in agricultural and veterinary sciences (13.3), while the minimum is in physics (2.5).

Proceeding to analysis at more detailed levels, the mean difference between % rank of top scientists and the rest of the population for the different indicators was calculated for the individual SDS of each discipline. The results are presented in Table 4: each cell indicates the number of SDS in which top scientists (by productivity) show an average quality that is higher than that of their colleagues in the same SDS. The top scientists show a scientific production with average quality greater than the rest of the population in 155 of the 165 SDS for $QI_c^{ave}$, in 160 SDS for $QI_c^{perc}$ and in 133 SDS for $QI_{if}$.

[Table 4]

*3.3 Analysis for best publications*

In the preceding sections we have shown that there is a significant correlation between productivity and average quality of output. In this section we will test for possible correlation between productivity and the quality of the best publication achieved by each researcher. For this, the regression analysis seen in previous section was repeated, with the objective of estimating the γ parameter for a power-law function of type $C_o = aP^\gamma$, in which in this case $C_o$ represents the maximum value of $QI_c^{ave}$ observed for the publications of a given researcher and P is his or her productivity. We can expect that, with an increasing number of publications achieved by an author, the normalized number of citations for the best one will increase less than proportionally. The results for each UDA, presented in Table 5, confirm the expectations.

[Table 5]

The table shows that be value for γ is positive and statistically significant, and also varies little among UDA. Thus the quality of the best publication by a scientist increases less than proportionally with respect to productivity. For example, in the physics discipline, the relative quality of the best publication of a given scientist is double that of a colleague who has a productivity of one third as much.

We now ask if it is the case that in comparing top scientists and the rest of the population, limiting the analysis to the best publications of each scientist, the correlation between productivity and quality which we see in the previous sections emerges as confirmed and further reinforced. The answer is definitely yes: the best publications of top scientists identified on the basis of P show a % rank that is higher by a mean of 34.5 for $QI_c^{ave}$, 34.8 for $QI_c^{perc}$ and 30.9 for $QI_{if}$, compared to the rest of the population (Table 6). At the level of the single SDS, the average quality of the best publications by top scientists is superior to that of the rest of the population in 159 of the total of 165

SDS: 6 SDS offer the exception, but only when the top scientists are identified on the basis of FP and quality is evaluated in terms of $QI_c^{ave}$.

[Table 6]

Still referring to the best publication of each researcher, a last type of comparison between the two subpopulations was conducted using the casual variables sequence criterion. Beginning with the impact ranking of the best publication of each top scientist in his/her SDS, the distance between the ideal and real cases was measured:

$$R^{diff}_{TS-j} = R^{max}_{TS-j} - R^{eff}_{TS-j}$$

where:

$R^{max}_{TS-j}$ = sum of the ranks of top scientists in sector j under the hypothesis of maximum differentiation (i.e. the situation in which the highest ranking non-top scientist is still ranked below the lowest top scientist).

$R^{eff}_{TS-j}$ = sum of the rank of all top scientists in the sector j

The value $R^{diff}_{TS-j}$ therefore represents the "distance" for the ideal situation of maximum quality difference between scientists in favor of top scientists. The same calculation is completed for non-top scientists ("others") and comparison of $R^{diff}_{TS-j}$ and $R^{diff}_{others-j}$ then identifies which of the two populations, top scientist or others, obtains a higher overall ranking. Conducting the comparison for each SDS of each UDA permits observation of how many SDS present the situation in which the average quality of publication by top scientists is less than that of the rest of the population. The results of this type of analysis for combinations of indicator of quality (of the publications) and of productivity (in the identification of top scientists) are presented in Table 7: columns 3 to 8 indicate the number of SDS in which top scientists presented an average quality rank that was less than that of their "non-top" colleagues, according to the criteria just given.

It can be observed that the presence of SDSs in which this event occurs is concentrated under two pairings of indicators: P- $QI_{if}$ (18) and FP- $QI_{if}$ (23). In addition, the industrial and information engineering discipline has the highest number of SDS (8 and 7, for these two pairings of indicators) in which the observations are different from the general rule of superior quality of publications by most productive researchers. Examining the others indicators of quality ($QI_c^{ave}$ and $QI_c^{perc}$), such exceptions occur in a truly limited number of SDS.

[Table 7]

## 4. Discussion and conclusions

The research question that stimulated this work concerned the possible existence of a trade-off between productivity of research work and quality of scientific results. The question has delicate implications for decision-makers' choices in identifying suitable systems of performance assessment and reward. The risk of "publication inflation" is seen as a key concern in performance assessment systems that have publication counts as a key criterion (Geuna and Martin, 2003), with the Australian experience noted as an illustrative lesson. Here, universities were awarded significant funds on the basis of aggregate publication counts, and distributed the funds internally, all with little attention to quality of output. Thus, Australian universities saw an increase in publication productivity between 1990 and 1998 but also a corresponding drop in relative quality (Butler, 2003).

In fact, the great majority of studies and evaluation exercises at the national and international level have emphasized the quality dimension of research output, while neglecting the productivity side of the activity. This is the case of the Italian VTR, which is primarily based on the evaluation of quality for a very limited sample of research products (less than 10% of the total per university). Abramo et al. (2009b) identify several criticisms for this type of system, showing that universities indicated as "top quality" are not necessarily those that are most productive. There is also a further hidden risk, just as important as inflating quantity at the expense of quality: in the long term, favoring quality over productivity could lead scientists to concentrate only on research that attempts to leap ahead, with the hope of achieving notable quality, but exposed to greater risks of failure. Other scientists, perhaps less outstanding, or with less resources, could be induced to reduce research activity and turn to other aspects of their role.

The analysis reported here actually demonstrates the existence of a strong correlation between quantity and quality of research production: scientists that are more productive in terms of quantity also achieve higher levels for quality in their research products. The difference between the quality of publications by "top" scientists, identified on the basis of their productivity, and the quality of publications by their colleagues in the same disciplines is very substantial and essentially constant under different measures and among various disciplines. The extremely large dataset and the highly detailed levels of analysis (single researchers and single sectors of disciplines) leave little doubt in interpreting and generalizing from the results.

The study, with reflection on the specific Italian context, may permit useful considerations concerning indications for policy. In Italy, comparative research evaluation is only beginning, and is still not linked to resource allocation. Universities and other public research institutes achieve access to economic and human resources through mechanisms that involve very little direct competition. Only recently have pressure from public opinion and the necessity to reduce and optimize expenses pushed policy makers to begin reworking mechanisms towards "reward" systems for research funding and advancement of university professionals. Such mechanisms should obviously be correlated to the strategic objectives for the system and its single components. In fact, reaching and maintaining excellence in science is seen as a necessity for long-term objectives of national socio-economic returns. However, aside from excellence, attention must also be given to the dimension of efficiency in the system, as it drives towards new knowledge. In the Italian case, the limited pressure on the actors in the system, given the current lack of adequate incentive programs, leaves researchers free to decide between "producing more" or "producing better". The current study shows that, under these conditions, the situation is one where the most brilliant scientists have succeeded in the dimensions of both productivity and quality. It could actually be that the very lack of incentive systems for either dimension has favored this correlation between productivity and quality, encouraging development of scientists that are both "efficient" and "excellent".

But this still leaves the problem of choosing which dimension to favor in incentive programs and broad strategies for scientific production, both for policy-makers, aiming for major socio-economic returns, and for research managers, aiming for the increased institutional prestige and access to better and substantial resources. The hypothesized trade-off between productivity and quality emerges reshaped by the results of the current analysis: it is very likely that the dilemma under discussion can best be resolved by adopting incentive systems that do not go to extremes in favoring one dimension over the other, but rather consider both. It is certainly clear that incentive systems are necessary, in suitable forms, to stimulate both higher average productivity and quality in national research systems.

| UDA | Obs | Correlation | $\gamma$ | correct $R^2$ |
|---|---|---|---|---|
| Mathematics and computer sciences | 1,612 | 0.751 | 1.169*** (0.026) | 0.564 |
| Physics | 1,736 | 0.876 | 1.266*** (0.020) | 0.768 |
| Chemistry | 2,514 | 0.895 | 1.275*** (0.017) | 0.801 |
| Earth sciences | 765 | 0.832 | 1.248*** (0.030) | 0.692 |
| Biology | 3,474 | 0.861 | 1.260*** (0.015) | 0.741 |
| Medicine | 6,027 | 0.880 | 1.242*** (0.010) | 0.774 |
| Agricultural and veterinary sciences | 1,446 | 0.800 | 1.212*** (0.024) | 0.640 |
| Civil engineering and architecture | 451 | 0.814 | 1.222*** (0.039) | 0.661 |
| Industrial and information engineering | 2,425 | 0.820 | 1.214*** (0.017) | 0.673 |
| Total | 20,450 | 0.865 | 1.248*** (0.005) | 0.748 |

*Table 1: Statistics regarding correlation of normalized citations ($lgC_0$) with productivity ($lgP$)*
*Dependent variable: $lgC_0$; OLS estimation method; robust standard errors shown in brackets. Statistical significance: \*p-value <0.10, \*\*p-value <0.05, \*\*\*p-value <0.01. "Obs" equals the number of researchers with at least one citation in the period examined.*

| | | Quality index | | |
|---|---|---|---|---|
| | | $QI_c^{ave}$ | $QI_c^{perc}$ | $QI_{if}$ |
| Productivity index for top scientists selection | P | 15.4 | 20.6 | 7.5 |
| | FP | 12.9 | 17.8 | 6.0 |

*Table 2: Mean difference in % rank between top scientists and the rest of the population, for each indicator of quality*

| UDA | $QI_c^{ave}$ | $QI_c^{perc}$ | $QI_{if}$ |
|---|---|---|---|
| Mathematics and computer sciences | 14.4 | 21.8 | 4.4 |
| Physics | 17.0 | 21.8 | 2.5 |
| Chemistry | 15.9 | 18.4 | 9.1 |
| Earth sciences | 13.4 | 20.1 | 3.1 |
| Biology | 15.0 | 19.2 | 6.7 |
| Medicine | 15.3 | 20.8 | 9.8 |
| Agricultural and veterinary sciences | 19.3 | 23.4 | 13.3 |
| Civil engineering and architecture | 17.1 | 26.2 | 7.5 |
| Industrial and information engineering | 14.6 | 21.1 | 5.2 |
| Total | 15.4 | 20.6 | 7.5 |

*Table 3: Mean difference in % rank for quality between top scientist (identified on the basis of P) and the rest of the population, by UDA*

| UDA | Number of SDS | $QI_c^{ave}$ | $QI_c^{perc}$ | $QI_{if}$ |
|---|---|---|---|---|
| Mathematics and computer sciences | 9 | 9 | 9 | 5 |
| Physics | 7 | 7 | 7 | 5 |
| Chemistry | 11 | 11 | 11 | 11 |
| Earth sciences | 12 | 12 | 12 | 8 |
| Biology | 19 | 19 | 19 | 17 |
| Medicine | 41 | 37 | 38 | 38 |
| Agricultural and veterinary sciences | 25 | 20 | 24 | 20 |
| Civil engineering and architecture | 5 | 5 | 5 | 5 |
| Industrial and information engineering | 36 | 35 | 35 | 24 |
| Total | 165 | 155 | 160 | 133 |

*Table 4: Number of SDS in which average % rank of top scientists for indicators of quality is greater than the rest of the population (top scientists being identified on the basis of productivity)*

| UDA | Obs | Correlation | γ | correct $R^2$ |
|---|---|---|---|---|
| Mathematics and computer sciences | 1,612 | 0.546 | 0.618*** (0.024) | 0.297 |
| Physics | 1,736 | 0.660 | 0.661*** (0.021) | 0.435 |
| Chemistry | 2,514 | 0.651 | 0.600*** (0.018) | 0.424 |
| Earth sciences | 765 | 0.622 | 0.649*** (0.030) | 0.386 |
| Biology | 3,474 | 0.629 | 0.642*** (0.015) | 0.396 |
| Medicine | 6,027 | 0.686 | 0.664*** (0.010) | 0.470 |
| Agricultural and veterinary sciences | 1,446 | 0.589 | 0.625*** (0.023) | 0.346 |
| Civil engineering and Architecture | 451 | 0.622 | 0.645*** (0.039) | 0.385 |
| Industrial and information engineering | 2,425 | 0.622 | 0.648*** (0.016) | 0.387 |
| Total | 20,450 | 0.657 | 0.640*** (0.006) | 0.431 |

*Table 5: Statistics regarding correlation of citations ($lgC_c$) of best publication of each single scientist with his/her productivity (lgP)*
*Dependent variable: $lgC_c$; OLS estimation method; robust standard errors in brackets. Statistical significance: \*p-value <0.10, \*\*p-value <0.05, \*\*\*p-value <0.01. "Observed" equals the number of researchers with at least one citation in the period examined.*

|  |  | Quality index | | |
|---|---|---|---|---|
|  |  | $QI_c^{ave}$ | $QI_c^{perc}$ | $QI_{if}$ |
| Productivity index for top scientists selection | P | 34.5 | 34.8 | 30.9 |
|  | FP | 31.2 | 31.7 | 28.3 |

*Table 6: Mean difference in % rank, for each indicator of quality, between the best publications of top scientists compared to those of the rest of the population*

| UDA | Total SDS | P-$QI_c^{perc}$ | P-$QI_c^{ave}$ | P-$QI_{if}$ | FP-$QI_c^{perc}$ | FP-$QI_c^{ave}$ | FP-$QI_{if}$ |
|---|---|---|---|---|---|---|---|
| Mathematics and computer sciences | 9 | 0 | 0 | 0 | 0 | 0 | 0 |
| Physics | 7 | 0 | 0 | 1 | 0 | 0 | 1 |
| Chemistry | 11 | 0 | 0 | 3 | 0 | 0 | 3 |
| Earth sciences | 12 | 0 | 0 | 1 | 0 | 0 | 3 |
| Biology | 19 | 0 | 0 | 1 | 0 | 0 | 3 |
| Medicine | 41 | 0 | 0 | 1 | 0 | 0 | 0 |
| Agricultural and veterinary sciences | 25 | 0 | 0 | 3 | 0 | 0 | 5 |
| Civil engineering and architecture | 5 | 0 | 0 | 0 | 0 | 0 | 1 |
| Industrial and information engineering | 36 | 0 | 1 | 8 | 3 | 3 | 7 |
| Total | 165 | 0 | 1 | 18 | 3 | 3 | 23 |

*Table 7: Number of SDS in which, according to the casual variables sequence criterion, the average quality of publication by top scientists is less than that for the rest of the population, for different combinations of indicators of quality and productivity.*